# Design of a Scalable Hybrid MAC Protocol for Heterogeneous M2M Networks

Yi Liu, Chau Yuen, *Senior Member*, *IEEE*, Xianghui Cao, *Member*, *IEEE*, Naveed Ul Hassan, and Jiming Chen, *Senior Member*, *IEEE*

*Abstract*—A robust and resilient medium access control (MAC) protocol is crucial for numerous machine-type devices to concurrently access the channel in a machine-to-machine (M2M) network. Simplex (reservation- or contention-based) MAC protocols are studied in most literatures which may not be able to provide a scalable solution for M2M networks with large number of heterogeneous devices. In this paper, a scalable hybrid MAC protocol, which consists of a contention period and a transmission period, is designed for heterogeneous M2M networks. In this protocol, different devices with preset priorities (hierarchical contending probabilities) first contend the transmission opportunities following the convention-based $p$-persistent carrier sense multiple access (CSMA) mechanism. Only the successful devices will be assigned a time slot for transmission following the reservation-based time-division multiple access (TDMA) mechanism. If the devices failed in contention at previous frame, to ensure the fairness among all devices, their contending priorities will be raised by increasing their contending probabilities at the next frame. To balance the tradeoff between the contention and transmission period in each frame, an optimization problem is formulated to maximize the channel utility by finding the key design parameters: the contention duration, initial contending probability, and the incremental indicator. Analytical and simulation results demonstrate the effectiveness of the proposed hybrid MAC protocol.

## I. Introduction

INTERNET of Things (IoT) is an integrated part of future internet including existing and evolving internet and network developments and could be conceptually defined as a dynamic global network infrastructure with self-configuring capabilities [1]–[3]. Machine-to-machine (M2M) represents a future IoT where billions to trillions of everyday objects and the surrounding environment are connected and managed through a range of devices, communication networks, and cloud-based servers. M2M communication is defined as the information exchange between machines and machines without any human interaction. M2M network is expected to be widely utilized in many fields of pervasive IoT applications [4]–[9], including industrial and agricultural automations, health care, transport systems, and electricity grids. Recently, the enormous economic benefits of the M2M communications drive intensive discussion in international standardization activities, such as local thermal equillibrium (LTE) and IEEE 802.16. There are several characteristics of M2M networks:

1) massive number of devices in service coverage and concurrent network access attempt from these devices;
2) high level of system automation in which the devices and systems can exchange and share data;
3) heterogeneous quality of service (QoS) in M2M network that may require priority-based medium access control (MAC) protocol;
4) fairness concern for different devices to share or compete the limited resources;
5) the battery-powered devices may need energy-efficient access and transmission control mechanism;
6) the data packet may be dropped due to time sensitivity of the data readings, infrequency, and small burst transmission in low deploy cost M2M network [10].

In order to handle the massive access in M2M, 3GPP LTE has several work items defined on M2M communications, primarily with respect to overload control [14], [15]. IEEE 802.16p proposals addressed enhancements for IEEE 802.16 m [16], [17] standard to support M2M applications. It is noted that the massive access management of M2M communication over wireless channels generally happen at the MAC layer. Hence, the design of a smart and efficient MAC protocol remains a key requirement for successful deployment of any M2M networks. As discussed by 3GPP and IEEE 802.16, the MAC protocol for M2M networks focused on contention-based random access (RA) schemes [19]–[21] that allow all of the devices contend the transmission opportunities in entire frame. The contention-based RA is popular due to its simplicity, flexibility, and low overhead. Devices can dynamically join or leave without extra operations. However, the transmission collisions are eminent when huge number of M2M devices tries to communicate the base station (BS) all at once. One of the solutions is to use reservation-based schemes such as time-division multiple access (TDMA) in M2M network. The TDMA scheme is well known as the collision-free access scheme where the transmission time

Manuscript received October 31, 2013; accepted February 27, 2014. Date of publication March 11, 2014; date of current version May 05, 2014. This work was supported in part by programs of NSFC under Grant 61370159, Grant U1035001, Grant U1201253, Grant 61203117, and Grant 61203036, in part by the Singapore University Technology and Design under Grant SUTD-ZJU/RES/02/2011, in part by National Program for Special Support of Top-Notch Young Professionals under Grant NCET-11-0445, and in part by Lahore University of Management Sciences (LUMS) via Faculty start-up Grant.
Y. Liu is with the Guangdong University of Technology, Guangzhou 510006, China, and also with the Singapore University of Technology and Design, Singapore 138682 (e-mail: yiliu115@gmail.com).
C. Yuen is with the Singapore University of Technology and Design, Singapore 138682 (e-mail: yuenchau@sutd.edu.sg).
N. U. Hassan is with the Department of Electrical Engineering, SSE, Lahore University of Management Sciences (LUMS), Lahore 54792, Pakistan (e-mail: naveed.hassan@yahoo.com).
X. Cao and J. Chen are with the State Key Laboratory of Industrial Control Technology, Department of Control, Zhejiang University, Hangzhou 310027, China (e-mail: xhcao@iipc.zju.edu.cn; jmchen@ieee.org).
Color versions of one or more of the figures in this paper are available online at http://ieeexplore.ieee.org.
Digital Object Identifier 10.1109/JIOT.2014.2310425

is divided into slots and each device transmits only during its own time slots [22]. The main defect of TDMA is the low transmission slot usage if only a small portion of devices have data to transmit.

Hence, the pure contention-based or reservation-based scheme may not be suitable to build up a scalable, flexible, and automatic communication structure for a dense heterogeneous M2M network. In [23] and [24], the researchers introduced a hybrid scheme which attempt to combine the best features of both of reservation-based and contention-based while offsetting their weaknesses. In [23], the authors designed the hybrid MAC scheme, for sensor network to adapt to the level of contention in the network under low contention; it behaves like carrier sense multiple access (CSMA), and under high contention, like TDMA. In [24], the authors proposed how to use hybrid MAC protocols to support video streaming over wireless networks. Such schemes tried to adapt to different bandwidth conditions depending on demand. In addition, some existing standards, such as IEEE 802.15.3, IEEE 802.15.4, and IEEE 802.11ad [25]–[27], adopt the hybrid MAC protocols to satisfy the advance wireless networking throughput. However, few of the existing researches and standards of the hybrid MAC protocol consider the heterogeneous applications for different users which may exist in M2M networks. In addition, considering the particular characteristics of the M2M network such as the massive access control, energy efficiency, and fairness, a new hybrid MAC protocol should be designed and evaluated.

In this paper, we first develop a hybrid MAC protocol for heterogeneous M2M networks, which will combine the benefit of both contention-based and reservation-based protocols. In this protocol, the contention and reservation process of different devices is a frame procedure which composes of two portions: contention only period (COP) and transmission only period (TOP). The COP is based on $p$-persistent CSMA mechanism which allows different devices to contend transmission slots with their own priorities, i.e., the contending probabilities. Only successful contending devices are allowed to transmit data during TOP that provides TDMA type of data communication. To ensure the fairness, if the devices failed in contention at the previous frame, their contending priorities will be raised by increasing the contending probabilities at the next frame. Given the frame duration, it is expected that the number of successful devices increases when the COP duration is prolonged. However, the COP duration increases at the expense of shortening the TOP, which results in the decrease of transmission slots. To achieve the optimal tradeoff between the contention and transmission period in each frame, an optimization problem is formulated to maximize the channel utility by deciding the optimal contending probability during COP, and the optimal number of devices allowed transmitting during TOP (which is related to the duration of COP). The analytical and simulation results demonstrate the effectiveness of the proposed hybrid MAC protocol.

In summary, we make the following contributions in this paper:
1) We design a scalable hybrid MAC protocol incorporating with $p$-persistent CSMA mechanism and TDMA mechanism for heterogeneous M2M networks.

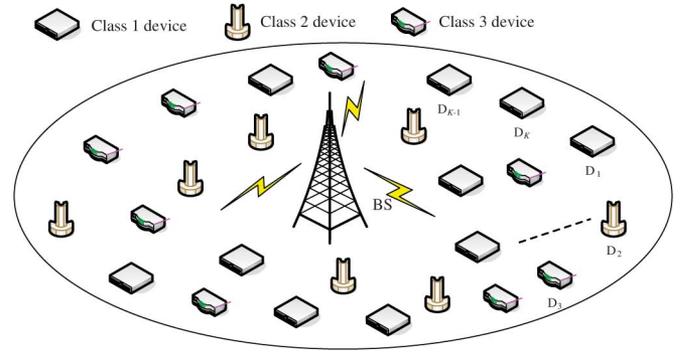

Fig. 1. System model of M2M network.

2) By defining different contending priorities, the protocol is able to allow the heterogeneous devices to obtain hierarchical performances.
3) We report an incremental contention priority method to guarantee the access fairness of devices, which failed to compete the transmission opportunities over time.
4) We identify the key design parameters to achieve maximum channel utility by considering the contending priority as well as the trade-off between contending period and transmission period.

The remainder of this paper is organized as follows. In Section II, we describe the system model of an M2M network. Then, we design a scalable hybrid MAC protocol in Section III. By optimizing the duration of COP and TOP, we introduce a hybrid access control scheme in Section IV. Performance study and evaluation are given in Section V. Section VI concludes the paper.

## II. M2M NETWORK MODEL

In this paper, we consider a heterogeneous M2M network, which consists of one BS and $K$ number of devices $\{D_1, \ldots, D_K\}$, as shown in Fig. 1. In this network, different types of devices are categorized into the $Q$ priority classes $\{C_1, \ldots, C_Q\}$, each of which has $K_q, \{q = 1, \ldots, Q\}$ number of devices and $K = \sum_{q=1}^{Q} K_q$. Without loss of generality, we assume that the higher classes of the devices have higher requirements of the transmission performance, such as higher channel utility and lower packet drop ratio, than that of the lower classes of devices. BS is responsible for operating MAC for different classes of the *active devices* (the devices that have packet to transmit) by assigning different contending probabilities $\{p_1, \ldots, p_Q\}$ to class $\{C_1, \ldots, C_Q\}$, respectively. We assume that the device with higher priority has higher contending probability than that with lower priority, i.e., $\{p_1 < p_2 \cdots < p_Q\}$.[1] To specifically describe the contending priorities of different classes of devices, we assume that the contending probabilities $p_q, \{q = 1, \ldots, Q\}$ have the relationship as

$$p_q = \max\{1, (1+\alpha)^{q-1} p_1\}, \quad 0 \leq p_1 \leq 1 \quad (1)$$

where $\alpha$ is defined as the incremental indicator.

For each active device, the data packet arrival process is modeled as a Possion arrival process with packet arrival rate $\lambda$.

---
[1]We can also set the priority to the packets instead of devices, e.g. some packets have higher priority than the others.

Here, for simplicity, we assume that all devices have the same packet arrival rate, which is known by BS. A new packet that arrives at a device is buffered until the device successfully contends the transmission opportunity and finishes the transmission. If a new packet arrives at the device before the transmission, the buffered packet will be replaced by the new one. Hence, there is one packet at most in the buffer of each device. Moreover, in order to maintain stationary property, we assume that all devices in the network are static (i.e., we do not consider mobility effects).

## III. A Scalable Hybrid MAC Protocol Design

In this section, we design a scalable hybrid MAC protocol for heterogeneous M2M network. We consider the operation of the M2M network on a frame-by-frame basis. Each frame is composed of four portions as depicted in Fig. 2: notification period (NP), COP, announcement period (AP), and TOP. During NP, the BS broadcasts notification message to all devices for notifying the beginning of the contention. The active devices will contend the channel during COP. The COP is based on $p$-persistent CSMA access method [29], and is used for devices to randomly send the transmission request to BS. After COP, the BS broadcasts the beginning of the transmission period during AP. The devices succeeded in contention is allowed to transmit data packet during the remaining time of a frame, which is specified as the TOP. The TOP provides a TDMA type of communication for the devices. We assume that each assigned transmission slot has the same length and there is no transmission error for each device [11], [12]. The specific description of the BS's and devices's operations in each period is given as follows.

### A. Operation of BS and Devices

*1) Notification Period:* At the start of every frame, the BS broadcasts a notification message to all $K$ number of devices to notify the beginning of the frame. Upon receiving the notification message, the active devices prepare to contend the transmission time slots. Other devices that do not have packets to send will enter sleep mode to preserve energy. By knowing packet arrival rate of every device, the BS estimates the number of active devices and calculates the optimal contending parameters: contending duration $T_{\text{COP,opt}}$, initial contending probability $p_{\text{inl,opt}}$, and incremental indicator $\alpha_{\text{opt}}$ by solving an optimization problem. The specific description and solution of the optimization problem is described in Section IV. These contending parameters are included in the notification message and broadcasted by BS during NP. Upon receiving the notification message, the active devices will calculate their own contending probabilities. The calculation is based on an incremental contending priority mechanism which will be presented in Section III-B. Next, the M2M network enters to COP.

*2) Contention Only Period:* In this period, we, respectively, present the operations of devices and BS as follows.

*Devices:* In this period, devices contend the transmission opportunities based on $p$-persistent CSMA mechanism, according to their own contending probability $p_\varrho$. The contending devices randomly send the transmission request (Tran-REQ) message to the BS. The contention is declared as success only when one device sends the Tran-REQ message. When more than one devices are sending Tran-REQ during the same time interval, a collision occurs. The idle period is a time interval in which the contention is not happening. Under $p$-persistent CSMA, the success period and collision period can be given as

$$\delta_{\text{coll}} = T_{\text{req}} + \text{BIFS}$$

and

$$\delta_{\text{succ}} = T_{\text{req}} + \text{SIFS} + T_{\text{ACK}} + \text{BIFS}$$

where $T_{\text{req}}$ is the length of Tran-REQ message, $T_{\text{ACK}}$ is the duration of acknowledge (ACK) message, and BIFS and SIFS are the backoff inter-frame space and short inter-frame space, respectively.

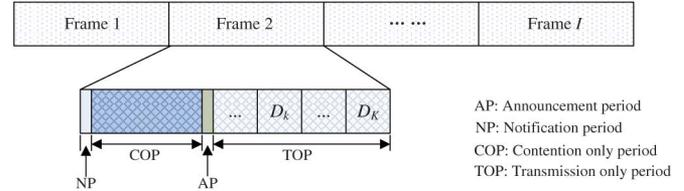

Fig. 2. Frame structure.

Upon receiving the ACK message from BS, the device will stop sending Tran-REQ message and waits for the AP duration. In addition, the ACK message includes the information of the index of the transmission time slot that the device is allowed to transmit in TOP.

*BS:* If a Tran-REQ message is successfully received from a device, the BS sends ACK message and the index of the transmission time slot to this device. For a given optimal contention period $T_{\text{COP,opt}}$, we may have more number of devices successfully contend a time slot than that is allowable for a given frame size (the expected number of successful devices during $T_{\text{COP,opt}}$ is denoted as $M_{\text{opt}}$). Hence, $T_{\text{COP,opt}}$ and $M_{\text{opt}}$ are used as two thresholds to control the duration of COP in practice. When the actual number of successful devices in contention is greater than $M_{\text{opt}}$ or the $T_{\text{COP}}$ is longer than $T_{\text{COP,opt}}$, the BS will stop the COP and declare the next period, i.e., AP.

*3) Announcement Period:* At the beginning of AP, BS initiates and broadcasts the announcement message to all of the devices. Upon receiving the announcement message, the devices succeeded in contention turn to transmission mode and prepare to send their own packets. Other active devices that failed in contention cease their contending operations and turn to sleep mode. Such arrangement keeps the wakeup time of a device at minimal, and we will further study the energy consumption of the proposed protocol in Section V.

*4) Transmission Only Period:* In TOP, active devices succeeded in contention sequentially operate the transmission following the TDMA mechanism. These devices turn ON their radio modules and transmit the data packet during their own transmission time slots. The devices turn OFF the radio module at other time slots. When the timer of TOP is out, the BS declares the beginning of a new frame. Although only uplink is mentioned, some modification to the protocol can be applied

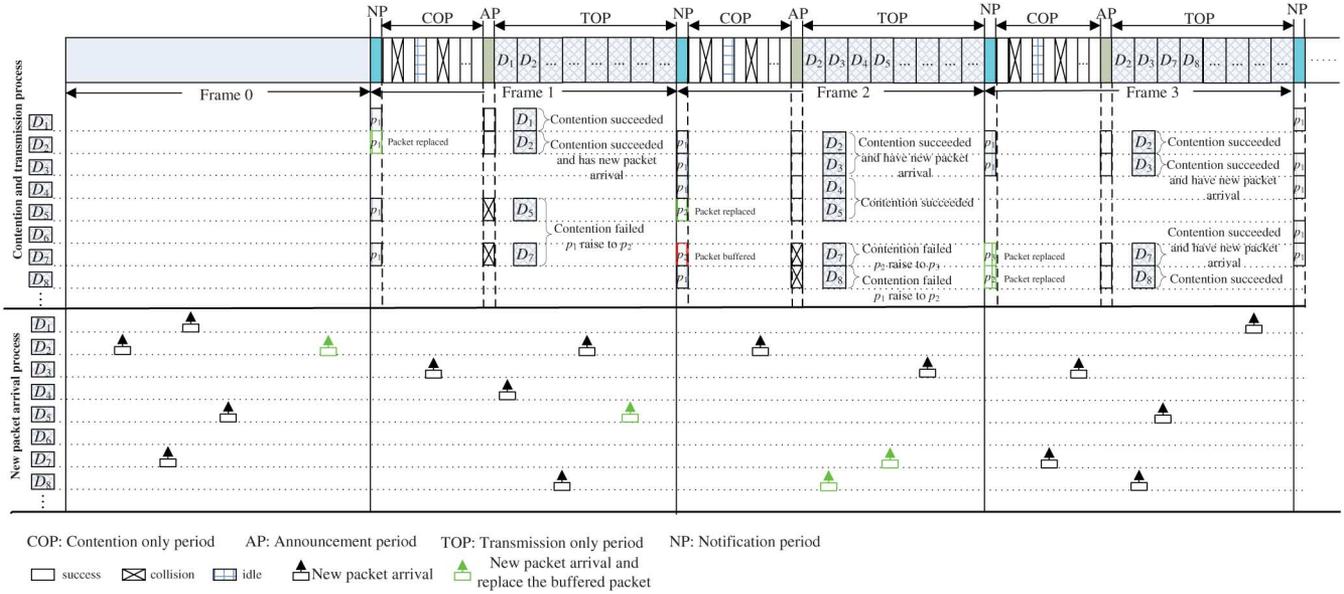

Fig. 3. Frame-based hybrid access process.

to downlink where the devices would like to receive information from the BS.

It is expected that higher number of devices succeeded in contention can be obtained if $T_{\text{COP}}$ gets longer. However, given the duration of a frame $T_{\text{frame}}$, the $T_{\text{TOP}}$ will be decreased as the $T_{\text{COP}}$ increases, which may reduce the total transmission time for allowing successful devices to transmit data. Hence, there is a tradeoff between the durations of COP and TOP. To balance this tradeoff, we intend to propose a hybrid access control scheme, presented in Section IV, which focuses on obtaining the optimal $T_{\text{COP}}$, $p_{\text{inl}}$, and $\alpha$ to maximize the channel utility.

### B. Incremental Contending Priority Mechanism for Fairness

According to $p$-persistent CSMA mechanism, the devices may fail in the contention during COP and lose the transmission opportunities at TOP. If a device frequently fails in contention frame-by-frame, the transmission performance of this device will drastically degrade. Hence, we propose an incremental contention priority model for the sake of fairness. In this model, we increase the contending probability of the device frame-by-frame if such device failed to contend the transmission opportunities at previous frames. When the device successfully transmit a data packet, the increasing process will be stopped and the priority of the device will return to the preliminary level. For type $q$ devices that have a new packet arrival, the BS assigns the preliminary contending probability $p_q$. If the type $q$ devices failed in contention at previous frames, the BS will increase their contending probability at current frame according to

$$p_{q,d} = \max\{1, (1+\alpha)^d p_q\}, \quad 0 \leq p_q \leq 1 \quad (2)$$

where $d = 0, 1, 2, \ldots$ is the number of frames during which the devices failed in contention and $\alpha$ is the incremental indicator. For simplicity, in this paper, we assume that the contending probability defined in (1) and (2) has the same incremental indicator $\alpha$ (in general, they can be two different parameters).

Hence, we note that there could be more than one class of devices that have the same contending probabilities at a certain frame. For example, there will be three kinds of devices that have the same contending probability at a frame: the class 3 devices have preliminary contending probability $p_{3,0} = (1+\alpha)^2 p_1$; the class 2 devices, which failed once in previous contention, have increased the contending probability to $p_{2,1} = (1+\alpha)p_2 = (1+\alpha)^2 p_1$; and the class 1 devices, which failed continuously across two frames, have increased the contending probability to $p_{1,2} = (1+\alpha)^2 p_1$. For easy expression, we define the *virtual class* $\varrho$, $\{\varrho = 1, \ldots, Q\}$ in which the devices, at a certain frame, have the same contending probability $p_\varrho$. Then, we have

$$p_\varrho = \max\{1, (1+\alpha)^\varrho p_{\text{inl}}\}, \quad 0 \leq p_{\text{inl}} \leq 1 \quad (3)$$

where $p_{\text{inl}} = p_1$ and $\varrho = q + d - 1, \{q = 1, \ldots, Q, d = 0, 1, 2, \ldots\}$.

### C. An Illustrative Example

An example of using the proposed protocol is illustrated in Fig. 3, where the hybrid access operations of eight devices are considered. To clearly describe our protocol, we only consider one type of priority class in the example, i.e., $q = 1$, then, $\varrho = d$, $\{d = 0, 1, 2, \ldots\}$. The operation of each device in the M2M network includes two processes: 1) contention and transmission process and 2) new packet arrival process. At frame 0, there is no contention and transmission process for any devices. We can see that the active devices are $D_1$, $D_2$, $D_5$, and $D_7$ at the end of frame 0. Note that during frame 0, $D_2$ has two packets arrival and the former arrived packet will be replaced by the latter one. At frame 1, $D_1$, $D_2$, $D_5$, and $D_7$ take part in contention with the contending probability $p_1$. $D_1$ and $D_2$, which succeeded in contention are allowed to transmit data during the following TOP. $D_5$ and $D_7$, which failed in contention, should wait for contention again in the next frame. Meantime, their contending probabilities increase from $p_1$ to $p_2$, where $p_2 = (1+\alpha)p_1$. In addition, we can see that $D_2$, $D_3$, $D_4$, $D_5$, and $D_8$ have new packet arrival in frame 1. Hence, the devices will contend in frame 2 are $D_2$, $D_3$, $D_4$, $D_5$, $D_7$, and $D_8$, where the contending probability of $D_5$ and $D_7$ is $p_2$ and that of the rest is $p_1$. After frame 2, we can see that $D_7$ failed in contention again and its contending probability $p_2$ increases to $p_3$, where $p_3 = (1 + \alpha)p_2$. Finally, $D_7$ succeed in frame 3 and also has a new packet arrival during frame 3. $D_7$ will go back to the contention at frame 4 with contending probability $p_1$.

## IV. Hybrid Multiple Access Control in Massive M2M Network

In this section, we first provide the expression of average duration of COP $\mathcal{T}_{\text{COP}}$ in terms of the number of successful devices $M$, initial contending probability $p_{\text{inl}}$, and incremental indicator $\alpha$. Then, we formulate an optimization problem to maximize the channel utility. The offline solution of the optimization problem will be given to find the optimal $T_{\text{COP,opt}}$, $p_{\text{inl,opt}}$, and $\alpha_{\text{opt}}$.

### A. Derivation of Average $T_{\text{COP}}$

According to our proposed MAC protocol, the active devices in virtual class $\varrho$ have the contending probability $p_\varrho$ and will contend the transmission slots during COP in a frame. To obtain the expression of $T_{\text{COP}}$, we need to know the average contending time of each device. First, we make the following assumptions and notations:

1) The packet arrival rate ($\lambda$) at each frame and the duration of each frame ($T_{\text{frame}}$) are constant.
2) For frame $i$, $\{i = 1, 2, \ldots, I\}$, there are $\Theta$ virtual classes of devices, each of which contains $N_\varrho^{(i)}$ devices ($1 \leq \varrho \leq \Theta$), where $N_\varrho^{(i)} = N_\varrho^{(i-1)} - M_\varrho^{(i-1)} + U_\varrho^{(i-1)}$, $M_{\varrho-1}^{(i-1)}$ is the number of $\varrho - 1$th class of devices that succeeded in contention at $(i-1)$th frame, $U_\varrho^{(i-1)}$ is the number of empty class $\varrho$ devices that have new packet arrival during $(i-1)$th frame.
3) A class $\varrho$ device uses the probability $p_\varrho = (1 + \alpha)^\varrho p_{\text{inl}}$ in the $p$-persistent CSMA ($0 \leq p_{\text{inl}} \leq 1, \alpha > 0$) to contend the transmission opportunities.

Based on $p$-persistent CSMA in COP period, when a contention attempt is completed (successfully or with a collision), the active device will start a contention attempt with probability $p_\varrho$, $\varrho = 1, \ldots, \Theta$. Here, we define the successful contention as the event that the transmission request from a device is successfully received by BS. Let $t_{m,i}$ denote the time between the $(m-1)$th and the $m$th successful contentions at frame $i$. Let $N_{m,i}^c$ denote the number of collisions that occur during $t_{m,i}$, then

$$t_{m,i} = \sum_{j=1}^{N_{m,i}^c} [\text{Idle}_{m,j} + \text{Coll}_{m,j}] + \text{Idle}_{N_{m,i}^c+1} + S_{m,i} \quad (4)$$

where $\text{Idle}_{m,j}$ is the duration of the $j$th idle time that precedes the channel busy period (either collision or success) in each $t_m$ duration. $\text{Coll}_{m,j}$ is the duration of the $j$th collision given that a collision occurs, and $S_{m,i}$ is the length of the request message. Let $E[t_i]$ denote the average contention time at frame $i$, then, we have

$$E[t_i] = (E[N_i^c] + 1)E[\text{Idle}_i] + E[N_i^c]E[\text{Coll}_i] + E[S_i] \quad (5)$$

where $E[N_i^c]$, $E[\text{Idle}_i]$, $E[\text{Coll}_i]$, and $E[S_i]$ are the average number of collisions, the average duration of an idle time, a collision, and a request message at frame $i$, respectively.

The total number of devices that can successfully contend the transmission opportunities at frame $i$ is denoted by $M^{(i)}$. Let $T_{\text{COP},i}$ denote the duration of the COP at steady frame $i$. Then, we have

$$T_{\text{COP},i} = \sum_{m=1}^{M^{(i)}} t_{m,i} \quad (6)$$

$$= \sum_{m=1}^{M^{(i)}} \left\{ \sum_{j=1}^{N_{m,i}^c} [\text{Idle}_{m,j} + \text{Coll}_{m,j}] + \text{Idle}_{N_{m,i}^c+1} + S_{m,i} \right\}. \quad (7)$$

Since $T_{\text{COP},i}$ is the sum of random variable $t_m$, ($m = 1, \ldots, M^{(i)}$), the $T_{\text{COP},i}$ is also a random variable with $E[T_{\text{COP},i}]$ the average time for $M^{(i)}$ number of successful contentions. To obtain the close-form expression of the $T_{\text{COP},i}$, we then focus on deriving the expected value of $T_{\text{COP},i}$, which is denoted by $E[T_{\text{COP},i}]$. Due to the independently distributed $t_{i,m}$, we have

$$\begin{aligned} E[T_{\text{COP},i}] &= M^{(i)} E[t_i] \\ &= M^{(i)} \{ (E[N_i^c] + 1)E[\text{Idle}_i] \\ &\quad + E[N_i^c]E[\text{Coll}_i] + E[S_i] \}. \end{aligned} \quad (8)$$

Then, we derive $E[N_i^c]$, $E[\text{Idle}_i]$, and $E[\text{Coll}_i]$ in the case of multiple priority classes.

*Derivation of $E[N_i^c]$*: Let $N_{\text{cd}}$ denote the number of contending devices in a contending slot immediately after an idle interval, and let $P_{\text{collision}}$ and $P_{\text{success}}$, respectively, denote the probability that a collision occurs and that a transmission is successful, both conditioned on that at least one device transmit the contending message. Then,

$$\begin{aligned} P_{\text{collision}} &= P(N_{\text{cd}} \geq 2 | N_{\text{cd}} \geq 1) \\ &= \frac{1 - P(N_{\text{cd}} = 0) - P(N_{\text{cd}} = 1)}{1 - P(N_{\text{cd}} = 0)} \\ &= 1 - \frac{\sum_{\varrho=1}^{\Theta} N_\varrho^{(i)} (1 - p_\varrho)^{N_\varrho^{(i)}-1} \prod_{\varrho \neq l} (1 - p_l)^{N_l^{(i)}}}{1 - \prod_{i=\varrho}^{\Theta} (1 - p_\varrho)^{N_\varrho^{(i)}}} \end{aligned} \quad (9)$$

and

$$\begin{aligned} P_{\text{success}} &= P(N_{\text{cd}} = 1 | N_{\text{cd}} \geq 1) \\ &= \frac{\sum_{i=\varrho}^{\Theta} N_\varrho^{(i)} (1 - p_\varrho)^{N_\varrho^{(i)}-1} \prod_{\varrho \neq l} (1 - p_l)^{N_l^{(i)}}}{1 - \prod_{\rho=1}^{\Theta} (1 - p_\varrho)^{N_\varrho^{(i)}}}. \end{aligned} \quad (10)$$

The probability distribution of $N_i^c$ can be expressed as

$$P(N_i^c = l) = P_{\text{collision}}^l P_{\text{success}} \quad (11)$$

and $E[N_i^c]$ can be obtained as

$$E[N_i^c] = \sum_{\varrho=1}^{\Theta} \varrho P(N_i^c = \varrho)$$

$$= \frac{1 - \prod_{\varrho=1}^{\Theta} (1-p_\varrho)^{N_\varrho^{(i)}}}{\sum_{\varrho=1}^{\Theta} N_\varrho^{(i)}(1-p_\varrho)^{N_\varrho^{(i)}-1} \prod_{\varrho \neq l}(1-p_l)^{N_l^{(i)}}} - 1. \quad (12)$$

*Derivation of* $E[\text{Idle}_i]$: Since class $\varrho$ devices may contend in a slot with probability $p_\varrho$, we have

$$E[\text{Idle}_i] = \delta_{\text{idle}} \sum_{\varrho=1}^{\Theta} \varrho P(N_{\text{cd}} \geq 0)(P(N_{\text{cd}} = 0))^\varrho$$

$$= \delta_{\text{idle}} \frac{\prod_{\varrho=1}^{\Theta} (1-p_\varrho)^{N_\varrho^{(i)}}}{1 - \prod_{\varrho=1}^{\Theta} (1-p_\varrho)^{N_\varrho^{(i)}}} \quad (13)$$

where $\delta_{\text{idle}}$ is a constant [29].

Let $\delta_{\text{coll}} = E[\text{Coll}_i]$ and $\delta_{\text{succ}} = E[S_i]$. Then, $E[T_{\text{COP},i}]$ is the function of $M^{(i)}$ and $p_\varrho$. Let $\mathcal{T}_{\text{COP}}^{(i)}(M^{(i)}, p_\varrho) = E[T_{\text{COP},i}]$, after some algebraic manipulations:

$$\mathcal{T}_{\text{COP}}^{(i)}(M^{(i)}, p_\varrho) =$$
$$M^{(i)} \left\{ \delta_{\text{idle}} \frac{\prod_{\varrho=1}^{\Theta} (1-p_\varrho)^{N_\varrho^{(i)}}}{\sum_{\varrho=1}^{\Theta} N_\varrho^{(i)}(1-p_\varrho)^{N_\varrho^{(i)}-1} \prod_{\varrho \neq l}(1-p_l)^{N_l^{(i)}}} \right.$$
$$\left. + \delta_{\text{coll}} \left( \frac{1 - \prod_{\varrho=1}^{\Theta} (1-p_\varrho)^{N_\varrho^{(i)}}}{\sum_{\varrho=1}^{\Theta} N_\varrho^{(i)}(1-p_\varrho)^{N_\varrho^{(i)}-1} \prod_{\varrho \neq l}(1-p_l)^{N_l^{(i)}}} - 1 \right) + \delta_{\text{succ}} \right\}. \quad (14)$$

In Section VI-B, we use $\mathcal{T}_{\text{COP}}^{(i)}(M^{(i)}, \alpha, p_{\text{inl}})$ to express $\mathcal{T}_{\text{COP}}^{(i)}(M^{(i)}, p_\varrho)$, since $p_\varrho = (1+\alpha)^\varrho p_{\text{inl}}$, where $\varrho = q+d-1$.

### B. Optimization Problem Formulation

Given $T_{\text{frame}}$, longer $\mathcal{T}_{\text{COP}}^{(i)}(M^{(i)}, \alpha, p_{\text{inl}})$ allows more devices to succeed in contention. However, the incremental $\mathcal{T}_{\text{COP}}^{(i)}(M^{(i)}, \alpha, p_{\text{inl}})$ will reduce the duration of TOP subjecting to the constraint as $\mathcal{T}_{\text{COP}}^{(i)}(M^{(i)}, \alpha, p_{\text{inl}}) + T_{\text{TOP}}^{(i)} \leq T_{\text{frame}}$. To balance this tradeoff, we formulate an optimization problem to maximize the channel utility in each frame. In this paper, the *channel utility*, denoted by $\mathcal{C}$, is defined as the mean value of the duration of TOP divided to the entire duration of frame $T_{\text{frame}}$ over $I$ frames

$$\mathcal{C} = \frac{1}{I} \sum_{i=1}^{I} \frac{T_{\text{TOP}}^{(i)}}{T_{\text{frame}}} = \frac{1}{I} \sum_{i=1}^{I} \frac{M^{(i)} T_r}{T_{\text{frame}}}$$

where $T_r$ denote the duration of a transmission time slot. Then, we can maximize the channel utility as follows:

$$\max_{\substack{\mathcal{T}_{\text{COP}}^{(i)}, i=1,\ldots,I \\ \alpha, p_{\text{inl}}}} \mathcal{C} \quad (15)$$

$$\text{s.t.} \quad \mathcal{T}_{\text{COP}}^{(i)}(M^{(i)}, \alpha, p_{\text{inl}}) + M^{(i)} T_r \leq T_{\text{frame}} \quad (16)$$
$$i = 1, \ldots, I$$
$$N_\varrho^{(i)} = N_{\varrho-1}^{(i-1)} - M_{\varrho-1}^{(i-1)} + U_\varrho^{(i-1)} \quad (17)$$
$$i = 1, \ldots, I$$
$$M_\varrho^{(i)} = \lceil M^{(i)} P_{\text{pck}}(\varrho) \rceil, \quad i = 1, \ldots, I \quad (18)$$
$$0 < p_{\text{inl}} \leq 1, \quad \alpha > 0. \quad (19)$$

The derivation of $\{U_\varrho^{(i-1)}, i = 1, \ldots, I\}$ in constraint (17) is given in Appendix A. In constraint (18), $P_{\text{pck}}(\varrho)$ is defined as the probability that a class $\varrho$ device successfully contend in the contending time

$$P_{\text{pck}}(\varrho) = \frac{N_\varrho^{(i)} p_\varrho (1-p_\varrho)^{N_\varrho^{(i)}-1} \prod_{\varrho \neq l}(1-p_l)^{N_l^{(i)}}}{\sum_{\varrho=1}^{\Theta} N_q^{(i)} p_\varrho (1-p_\varrho)^{N_\varrho^{(i)}-1} \prod_{\varrho \neq l}(1-p_l)^{N_l^{(i)}}}. \quad (20)$$

Hence, the number of class $\varrho$ devices successfully contend the transmission opportunities at frame $i$ is given by

$$M_\varrho^{(i)} = \lceil M^{(i)} P_{\text{pck}}(\varrho) \rceil.$$

Note that $\mathcal{T}_{\text{COP}}^{(i)}$ is the function of $M^{(i)}$ as shown in (14). For simplicity of expression, we use variable $M^{(i)}$ instead of $\mathcal{T}_{\text{COP}}^{(i)}$. Finally, the previous optimization problem can be written as

$$\max_{\substack{M^{(i)}, i=1,\ldots,I \\ \alpha, p_{\text{inl}}}} \frac{T_r}{I \cdot T_{\text{frame}}} \sum_{i=1}^{I} M^{(i)}$$
$$\text{s.t.} \quad \text{constraints} \quad (16)–(19). \quad (21)$$

Next, we try to prove the convexity of the above-mentioned optimization problem. It is easy to observe that the objective function (15) is a linear function of $M^{(i)}$ and constraints (17)–(19) are linear. For constraint (16), we provide the convexity proof by Theorem 1.

*Theorem 1:* Let $L = \sum_{\varrho=1}^{\Theta} N_\varrho^{(i)}$, for $L \to \infty$ and $(1+\alpha)p_{\text{inl}} \leq 1$, $\mathcal{T}_{\text{COP}}^{(i)}(M^{(i)}, \alpha, p_{\text{inl}})$ can be obtained as a convex function of $M^{(i)}$, $\alpha$, and $p_{\text{inl}}$.

*Proof:* The proof is presented in Appendix B.　■

Theorem 1 shows that, asymptotically, for M2M networks with tremendous number of devices, i.e., when the number of active devices is large, constraint (16) is also a convex function. Therefore, the optimization problem is a convex programming problem and can be solved easily with off-the-shelf toolbox and the optimal period of COP, $\mathcal{T}_{\text{COP,opt}}^{(i)} = \mathcal{T}_{\text{COP}}^{(i)}(M_{\text{opt}}^{(i)}, \alpha_{\text{opt}}, p_{\text{inl,opt}})$.

TABLE I
SIMULATION PARAMETERS

| | | |
|---|---|---|
| $T_{frame}$ | 1000 ms | The duration of a frame |
| $T_r$ | 2 ms | The transmission time of each device |
| $T_{req}$ | 22.2 μs | The length of Tran-REQ message |
| $T_{NOF}$ | 10 μs | The length of notification message |
| $T_{ANC}$ | 10 μs | The length of announcement message |
| $T_{ACK}$ | 7.5 μs | The duration of ACK frame |
| $SIFS$ | 2.5 μs | The duration of short interframe spacing |
| $BIFS$ | 7.5 μs | The duration of backoff interframe spacing |
| $\mathcal{P}_t$ | 1.5 W | The transmission power of a device |
| $\mathcal{P}_r$ | 1 W | The receiving power of a device |
| $\mathcal{P}_i$ | 0.5 W | The idle power of a device |

TABLE II
UTILITY IN TERMS OF $p_{inl}$ AND $\alpha$ WHEN $\lambda = 1$

(a) 500 devices case

| Utility | $p_{inl}$=0.1 | $p_{inl}$=0.2 | $p_{inl}$=0.3 | $p_{inl}$=0.4 | $p_{inl}$=0.5 | $p_{inl}$=0.6 | $p_{inl}$=0.7 | $p_{inl}$=0.8 | $p_{inl}$=0.9 | $p_{inl}$=1.0 |
|---|---|---|---|---|---|---|---|---|---|---|
| $\alpha$=0.5 | 0.6180 | 0.6220 | 0.6219 | 0.6210 | 0.6201 | 0.6181 | 0.6150 | 0.6106 | 0.6061 | 0.5989 |
| $\alpha$=0.6 | 0.6182 | 0.6221 | 0.6220 | 0.6213 | 0.6203 | 0.6182 | 0.6151 | 0.6110 | 0.6061 | 0.5991 |
| $\alpha$=0.7 | 0.6183 | 0.6223 | 0.6224 | 0.6215 | 0.6207 | 0.6183 | 0.6156 | 0.6111 | 0.6062 | 0.5993 |
| $\alpha$=0.8 | 0.6191 | 0.6223 | 0.6225 | 0.6215 | 0.6209 | 0.6185 | 0.6158 | 0.6112 | 0.6063 | 0.5994 |
| $\alpha$=0.9 | 0.6192 | 0.6224 | 0.6226 | 0.6216 | 0.6210 | 0.6185 | 0.6159 | 0.6114 | 0.6064 | 0.5994 |
| $\alpha$=1 | 0.6198 | 0.6227 | 0.6229 | 0.6223 | 0.6210 | 0.6188 | 0.6160 | 0.6121 | 0.6067 | 0.5996 |
| $\alpha$=2 | 0.6196 | 0.6226 | 0.6227 | 0.6221 | 0.6210 | 0.6186 | 0.6159 | 0.6119 | 0.6066 | 0.5995 |
| $\alpha$=3 | 0.6194 | 0.6225 | 0.6225 | 0.6220 | 0.6208 | 0.6184 | 0.6158 | 0.6118 | 0.6065 | 0.5995 |
| $\alpha$=4 | 0.6191 | 0.6222 | 0.6224 | 0.6219 | 0.6207 | 0.6183 | 0.6156 | 0.6117 | 0.6064 | 0.5994 |
| $\alpha$=5 | 0.6188 | 0.6220 | 0.6223 | 0.6217 | 0.6205 | 0.6182 | 0.6155 | 0.6115 | 0.6064 | 0.5993 |

(b) 800 devices case

| Utility | $p_{inl}$=0.1 | $p_{inl}$=0.2 | $p_{inl}$=0.3 | $p_{inl}$=0.4 | $p_{inl}$=0.5 | $p_{inl}$=0.6 | $p_{inl}$=0.7 | $p_{inl}$=0.8 | $p_{inl}$=0.9 | $p_{inl}$=1.0 |
|---|---|---|---|---|---|---|---|---|---|---|
| $\alpha$=0.5 | 0.6745 | 0.6710 | 0.6660 | 0.6552 | 0.6381 | 0.6181 | 0.5960 | 0.5651 | 0.5398 | 0.5055 |
| $\alpha$=0.6 | 0.6747 | 0.6712 | 0.6662 | 0.6552 | 0.6403 | 0.6188 | 0.5961 | 0.5652 | 0.5388 | 0.5057 |
| $\alpha$=0.7 | 0.6750 | 0.6723 | 0.6665 | 0.6553 | 0.6403 | 0.6190 | 0.5962 | 0.5659 | 0.5371 | 0.5056 |
| $\alpha$=0.8 | 0.6751 | 0.6729 | 0.6669 | 0.6555 | 0.6404 | 0.6191 | 0.5963 | 0.5660 | 0.5366 | 0.5058 |
| $\alpha$=0.9 | 0.6753 | 0.6731 | 0.6682 | 0.6558 | 0.6405 | 0.6192 | 0.5966 | 0.5663 | 0.5351 | 0.5060 |
| $\alpha$=1 | 0.6759 | 0.6760 | 0.6707 | 0.6610 | 0.6481 | 0.6202 | 0.5970 | 0.5664 | 0.5398 | 0.5068 |
| $\alpha$=2 | 0.6758 | 0.6749 | 0.6668 | 0.6542 | 0.6404 | 0.6189 | 0.5949 | 0.5664 | 0.5388 | 0.5062 |
| $\alpha$=3 | 0.6754 | 0.6732 | 0.6653 | 0.6547 | 0.6403 | 0.6191 | 0.5932 | 0.5662 | 0.5371 | 0.5053 |
| $\alpha$=4 | 0.6753 | 0.6717 | 0.6652 | 0.6548 | 0.6397 | 0.6188 | 0.5939 | 0.5654 | 0.5366 | 0.5051 |
| $\alpha$=5 | 0.6749 | 0.6716 | 0.6654 | 0.6541 | 0.6305 | 0.6179 | 0.5931 | 0.5645 | 0.5351 | 0.5041 |

(c) 1200 devices case

| Utility | $p_{inl}$=0.1 | $p_{inl}$=0.2 | $p_{inl}$=0.3 | $p_{inl}$=0.4 | $p_{inl}$=0.5 | $p_{inl}$=0.6 | $p_{inl}$=0.7 | $p_{inl}$=0.8 | $p_{inl}$=0.9 | $p_{inl}$=1.0 |
|---|---|---|---|---|---|---|---|---|---|---|
| $\alpha$=0.5 | 0.7862 | 0.7774 | 0.7672 | 0.7530 | 0.7178 | 0.6902 | 0.6762 | 0.5906 | 0.4846 | 0.2882 |
| $\alpha$=0.6 | 0.7866 | 0.7786 | 0.7672 | 0.7548 | 0.7189 | 0.6914 | 0.6650 | 0.5974 | 0.4866 | 0.2956 |
| $\alpha$=0.7 | 0.7870 | 0.7810 | 0.7684 | 0.7466 | 0.7194 | 0.6956 | 0.6632 | 0.6020 | 0.5038 | 0.3018 |
| $\alpha$=0.8 | 0.7878 | 0.7864 | 0.7698 | 0.7470 | 0.7210 | 0.6958 | 0.6536 | 0.6021 | 0.5086 | 0.3040 |
| $\alpha$=0.9 | 0.7882 | 0.7868 | 0.7778 | 0.7500 | 0.7250 | 0.6988 | 0.6674 | 0.6042 | 0.5122 | 0.3358 |
| $\alpha$=1 | 0.7888 | 0.7874 | 0.7782 | 0.7530 | 0.7280 | 0.7032 | 0.6762 | 0.6090 | 0.5146 | 0.3882 |
| $\alpha$=2 | 0.7878 | 0.7846 | 0.7692 | 0.7548 | 0.7202 | 0.7014 | 0.6650 | 0.6374 | 0.4866 | 0.3556 |
| $\alpha$=3 | 0.7870 | 0.7810 | 0.7684 | 0.7466 | 0.7196 | 0.6956 | 0.6632 | 0.6220 | 0.4438 | 0.3218 |
| $\alpha$=4 | 0.7866 | 0.7764 | 0.7698 | 0.7470 | 0.7110 | 0.6956 | 0.6536 | 0.5878 | 0.4286 | 0.3140 |
| $\alpha$=5 | 0.7836 | 0.7748 | 0.7678 | 0.7400 | 0.7050 | 0.6928 | 0.6674 | 0.6342 | 0.4088 | 0.3058 |

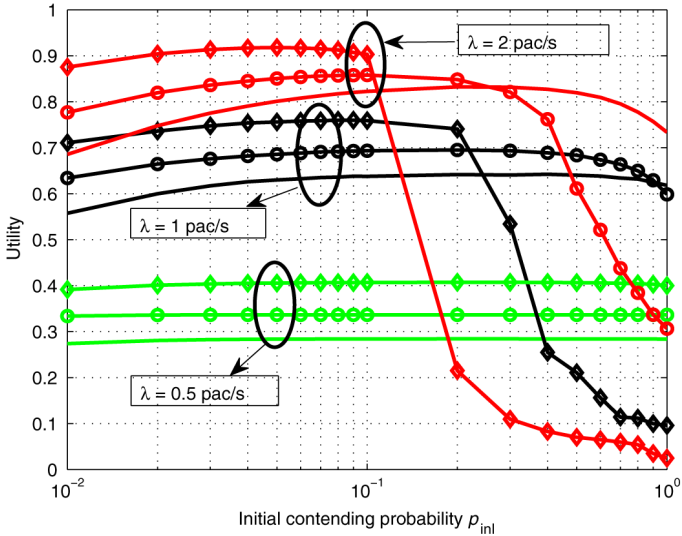

Fig. 4. Channel utility with different packet arrival rates in terms of $p_{inl}$ when $\alpha = 1$ in 500 (straight line), 800 (circle), and 1200 (diamond) cases.

## V. PERFORMANCE STUDY AND EVALUATION

In this section, we analyze channel utility, packet drop ratio, average transmission delay, and energy consumption of the proposed hybrid MAC protocol in heterogeneous M2M network, which consist of $K$ ($K = 500, 800, 1200$) number of devices. Without loss of generality, in our simulation, the heterogeneous M2M network includes three classes of devices: class 1 (device $D_{21} - D_K$), class 2 (device $D_1 - D_{10}$), and class 3 (device $D_{11} - D_{20}$), which have the contending probabilities $p_1 = p_{inl}, p_2 = (1+\alpha)p_{inl}$, and $p_3 = (1+\alpha)^2 p_{inl}$, respectively. Meanwhile, the performance comparisons of the proposed protocol with contention-based protocol—$p$-persistent CSMA [29] and reservation-based protocol—TDMA [22] are provided. Summarily, the simulation parameters are shown in Table I.

### A. Channel Utility

Fig. 4 shows the channel utility with different packet arrival rates, in terms of the initial contending probability $p_{inl}$ in 500 (straight line), 800 (circle), and 1200 (diamond) devices cases. It is observed that the channel utility increases at first and then decreases as the $p_{inl}$ increases. This is because the number of successful devices in contention increases as the contending probability increases for fixed total number of devices. Accordingly, the transmission of the successful devices will lead to increasing channel utility of a frame. However, as the contending probability increases to or exceeds a certain value, the number of contending collisions may become large and the duration of the contention may increase, which results in the decrement of transmission period (TOP).

Table II shows the channel utility in 500, 800, and 1200 devices cases when $\lambda = 1$, in terms of the initial contending probability $p_{inl}$, and incremental indicator $\alpha$. From Table II, we observe that the channel utility is maximized as 0.6229 ($\alpha_{opt} = 1$ and $p_{inl,opt} = 0.3$), 0.6760 ($\alpha_{opt} = 1$ and $p_{inl,opt} = 0.2$), and 0.7888 ($\alpha_{opt} = 1$ and $p_{inl,opt} = 0.1$) in 500, 800, and 1200 devices cases, respectively. The channel utility in the small number of devices case is lower than that in the large number of devices case. This is expected since the channel utilization may be reduced in the small number of devices condition due to limited transmission requirements through the entire network.

Fig. 5 shows the channel utility comparison in terms of packet arrival rate $\lambda$ among the proposed hybrid protocol, CSMA, and TDMA protocols in 1200 devices case. In this case, the proposed hybrid protocol chooses the optimal parameters for devices as $\alpha_{opt} = 1$ and $p_{inl,opt} = 0.1$. It is observed that the channel utility in the proposed hybrid protocol is first higher and then lower than TDMA, and will be higher than $p$-persistent CSMA as packet arrival rate $\lambda$ increases. That is, the proposed hybrid protocol can optimally control the initial contending probability $p_{inl,opt}$, incremental indicator $\alpha$, and the number of successful devices to maximize the channel utility. While $p$-persistent CSMA may perform well only at low-load condition, TDMA performs well only at heavy-load condition. Our hybrid protocol outperforms the TDMA when $\lambda$ is small and performs better than $p$-persistent CSMA when $\lambda$ is increased.

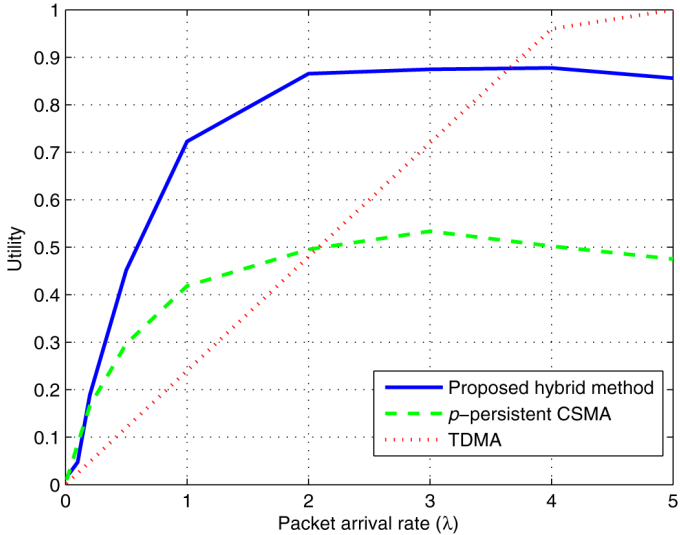

Fig. 5. Channel utility comparison in terms of packet arrival rate $\lambda$ in 1200 devices case when $\alpha_{\mathrm{opt}} = 1$ and $p_{\mathrm{inl,opt}} = 0.1$.

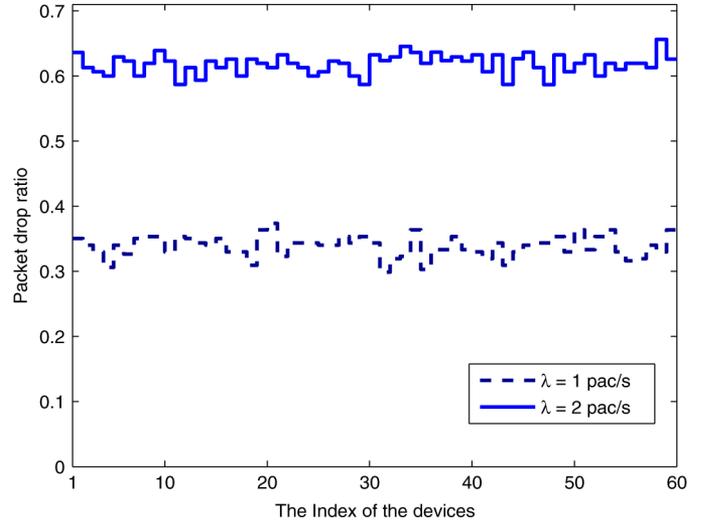

Fig. 7. Packet drop ratio with packet arrival rates (1 pac/s and 2 pac/s) in 1200 homogeneous devices case when $p_{\mathrm{inl,opt}} = 0.1$ and $\alpha_{\mathrm{opt}} = 1$.

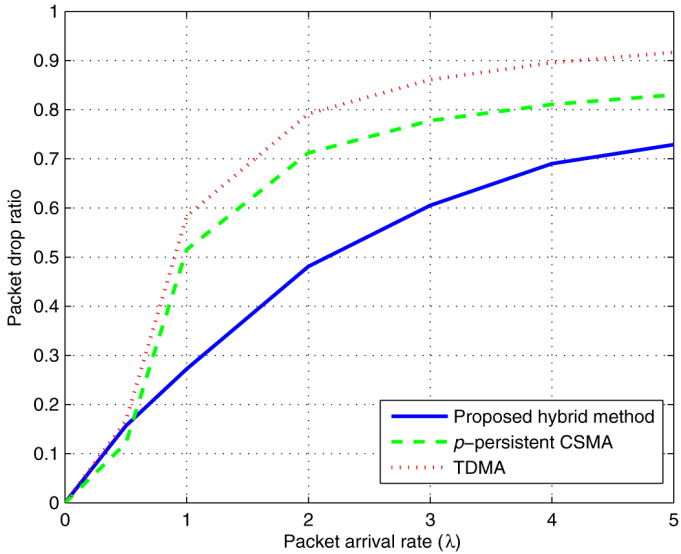

Fig. 6. Comparison of packet drop ratio in terms of packet arrival rate ($\lambda$) in 1200 devices case when $p_{\mathrm{inl,opt}} = 0.1$ and $\alpha_{\mathrm{opt}} = 1$.

### B. Packet Drop Ratio

In this paper, we use packet drop ratio as a metric to measure the network throughout. The packet drop ratio is defined as the ratio of the number of dropped packets to the total number of generated packets per device during the simulation duration. Fig. 6 shows the packet drop ratio of the proposed hybrid protocol in 1200 devices case when $p_{\mathrm{inl,opt}} = 0.1$ and $\alpha_{\mathrm{opt}} = 1$. We observe that the packet drop ratio of the proposed protocol is first higher and then lower than TDMA and $p$-persistent CSMA as packet arrival rate $\lambda$ increases. That is, the proposed hybrid protocol is able to increase the contending probability for the device when it continually failed in the contention in previous frames. This mechanism can guarantee that all devices have fair chance to obtain the transmission slot, which leads to low packet drop ratio. Without this mechanism, the TDMA and $p$-persistent CSMA may perform well only at low-load condition. When packet arrival rate is increasing, the packet drop ratio of both of TDMA and $p$-persistent CSMA will be drastically increasing.

In Section V-B-1, we focus on analyzing the packet drop ratio of the proposed hybrid protocol in both of the homogeneous and heterogeneous M2M networks, which consist of 1200 devices.

*1) Homogeneous Case:* Different from the setting in heterogeneous case, all devices have the same contending probability $p = 0.1$ in homogeneous case. Fig. 7 shows the packet drop ratio of the homogeneous network with $\lambda = 1$ pac/s and $\lambda = 2$ pac/s in 1200 homogeneous devices case when $p_{\mathrm{inl,opt}} = 0.1$ and $\alpha_{\mathrm{opt}} = 1$. We observe that the packet drop ratio of the devices is nearly the same in 1 pac/s case (between 0.3 and 0.4) and 2 pac/s case (between 0.55 and 0.65), respectively. That is, based on our protocol, the devices that failed in contention at current frame will be given higher contending priorities in the following frame. This adaptive adjustment can guarantee that all of the devices in M2M network have the fair opportunity to compete the transmission time slots.

*2) Heterogeneous Case:* Fig. 8 shows the packet drop ratio when packet arrival rate is $\lambda = 1$ pac/s and $\lambda = 2$ pac/s in heterogeneous 1200 devices case when $p_{\mathrm{inl,opt}} = 0.1$ and $\alpha_{\mathrm{opt}} = 1$. We observe that class 3 devices have the lowest packet drop ratio and class 1 devices have the highest packet drop ratio on an average. This indicates that the different contending probabilities $p$ have big influence on the performance of the different classes devices in heterogeneous M2M networks.

### C. Average Transmission Delay

In this section, we define the *average transmission delay* as the average number of frames during which the active device successfully finishes a transmission. Suppose that a device is activated (new packet arrival) at frame $k_1$ and successfully finishes its transmission at frame $k_2$, where $1 \leq k_1, \ldots, \leq k_2 \leq I$. Then, the average transmission delay of a device can be obtained by

$$\text{Delay} = \frac{\sum_{k_1, k_2 \in \{1, \ldots, I\}} k_2 - k_1}{W}$$

where $W$ is the number of successful transmissions of the device during simulation. In simulation, we run the operation of the M2M network over $I = 200$ frames.

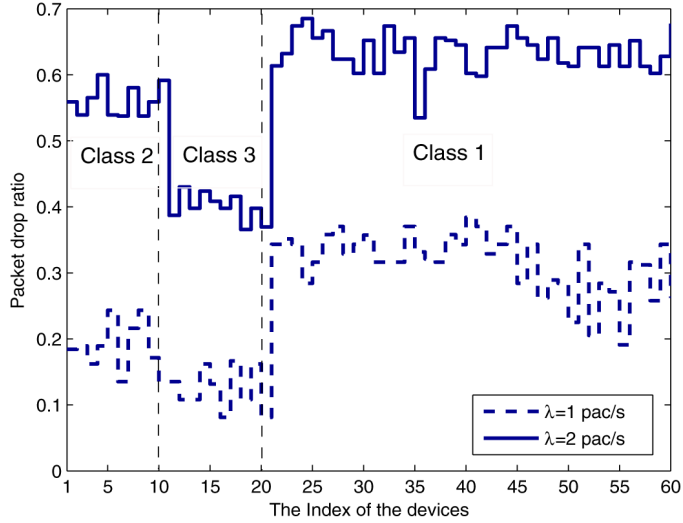

Fig. 8. Packet drop ratio with packet arrival rates (1 pac/s and 2 pac/s) in heterogeneous 1200 devices case when $p_{\text{inl,opt}} = 0.1$ and $\alpha_{\text{opt}} = 1$.

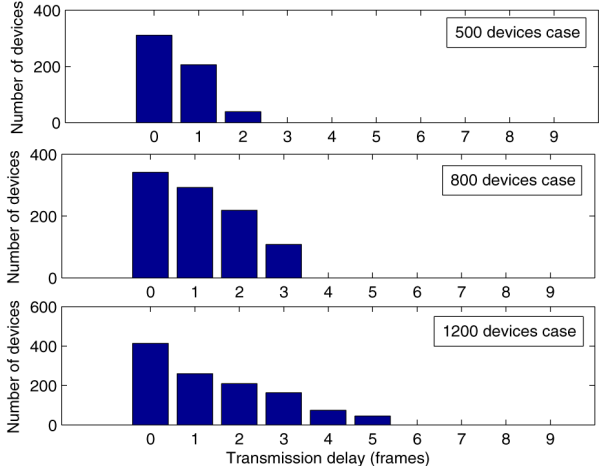

Fig. 9. Average transmission delay with packet arrival rate (1 pac/s) in 500, 800, and 1200 devices case when $p_{\text{inl}} = 0.1$ and $\alpha = 1$.

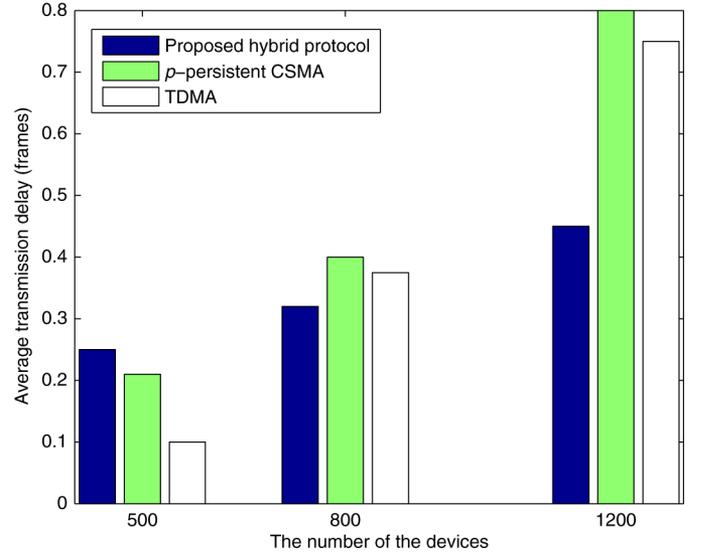

Fig. 10. Comparison of average transmission delay in terms of the total number of devices.

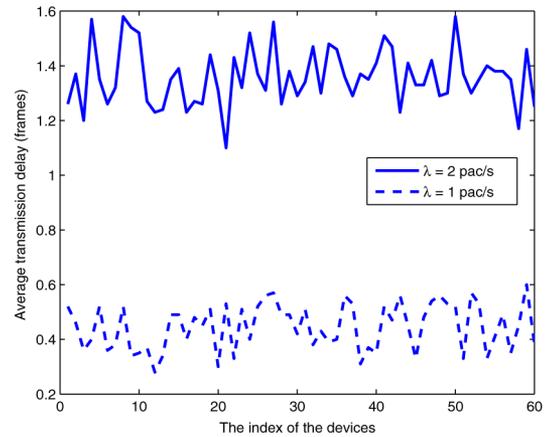

Fig. 11. Average transmission delay with packet arrival rates (1 pac/s and 2 pac/s) in 1200 devices case when $p_{\text{inl}} = 0.1$ and $\alpha = 1$.

The average transmission delay in 500, 800, and 1200 devices cases with packet arrival rate (1 pac/s) when $p_{\text{inl}} = 0.1$ and $\alpha = 1$ are shown in Fig. 9. We can observe that the average transmission delay increases as the number of devices increases. This is because the growing number of devices will cause the increasing collisions at COP of a frame, which leads to the increasing transmission delay.

Fig. 10 shows the comparison of average transmission delay among proposed hybrid method, $p$-persistent CSMA, and TDMA in 500, 800, and 1200 devices cases. In each case, the packet arrival $\lambda = 1$. For the proposed hybrid protocol, $p_{\text{inl}} = 0.1$ and $\alpha = 1$. The comparison indicates that the proposed hybrid protocol is able to obtain less transmission delay than $p$-persistent CSMA and TDMA in heavy load case (800 and 1200 devices cases). This is because the hybrid protocol can optimally control the contention period to allow more devices to have the transmission opportunities when the number of devices becomes very large.

Next, we also show the average transmission delay of the proposed hybrid protocol in both of the homogeneous and heterogeneous M2M networks, where the number of devices is 1200. Similar to the packet drop ratio evaluation, we first select 60 derives among the 1200 devices to show the simulation results.

*1) Homogeneous Case:* In homogeneous case, all devices have the same contending probability $p = 0.1$. Fig. 11 shows the average transmission delay of the devices 1–60 when the packet arrival rate of each devices are 1 and 2 pac/s, respectively. It is observed that the delay of each device fluctuate stably centered on 0.4 frames in 1 pac/s case and 1.6 frames in 2 pac/s case. This indicates that our proposed hybrid protocol can guarantee the fairness of the devices to obtain the transmission opportunities.

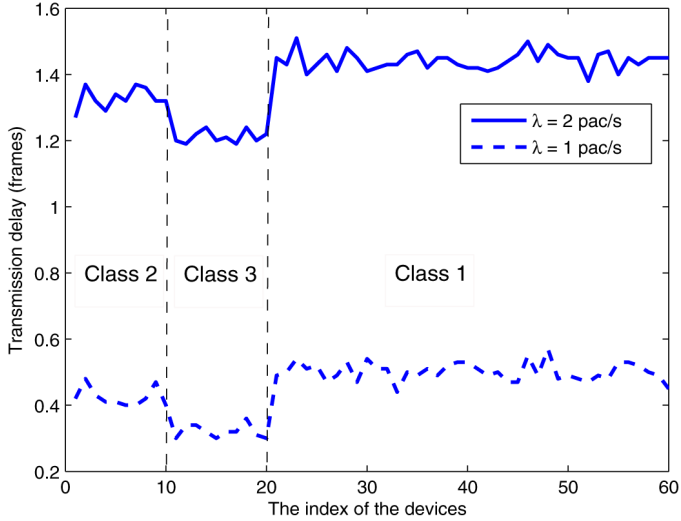

Fig. 12. Average transmission delay with packet arrival rates (1 pac/s and 2 pac/s) in heterogeneous 1200 devices case when $p_{\text{inl}} = 0.1$ and $\alpha = 1$.

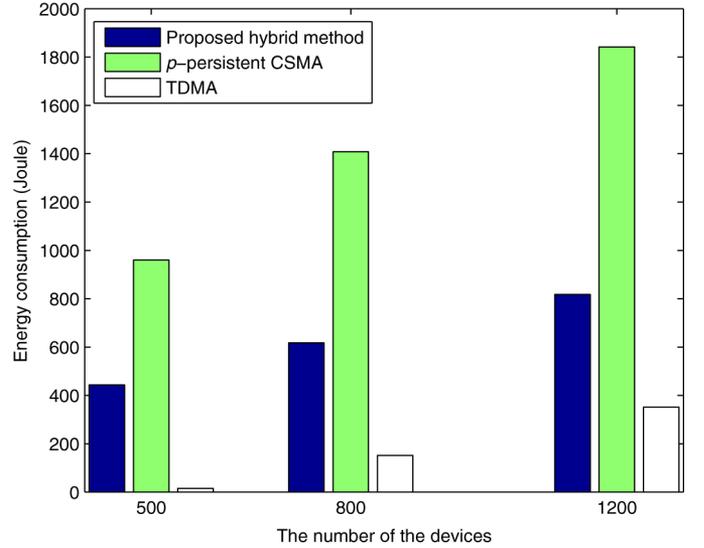

Fig. 13. Comparison of energy consumption in terms of the total number of devices.

*2) Heterogeneous Case:* Fig. 12 shows the average transmission delay of all 60 devices when the packet arrival rates of each device are 1 and 2 pac/s, respectively. As expected, it is shown that the class 3 devices have the lowest transmission delay and class 1 devices have the highest average delay.

### D. Energy Consumption

In this section, we consider the energy consumption of the M2M network in one frame. In this paper, we consider that BS obtains power from grid and the devices obtain power from battery. Hence, we focus on the power consumption of the devices. For a device, the power consumption during the transmission mode is denoted by $\mathcal{P}_t$ W; the power consumption during the receiving mode is denoted by $\mathcal{P}_r$ W; and the power consumption during the idle mode is denoted by $\mathcal{P}_i$ W. The energy consumption of the M2M network in each period is defined as follows.

During NP duration, each device receives a notification message. Let $E_{\text{NP}}$ denote the total energy used for receiving the notification messages during NP. Then, we have

$$E_{\text{NP}} = K\mathcal{P}_r T_{\text{NOF}}$$

where $K$ is the number of all devices in M2M network and $T_{\text{NOF}}$ is the length of the notification message.

During COP duration, there will be $M_{\text{opt}}$ devices that successfully send Tran-REQ message to BS. The total energy used for sending this message is denoted by $E_{\text{COP}}$, we have

$$E_{\text{COP}} = \sum_{m=1}^{M_{\text{opt}}} E_{m,i} \qquad (22)$$

where $E_{m,i} = \sum_{j=1}^{N_{m,i}^c} [\text{Idle}_{m,j}\mathcal{P}_i + \text{Coll}_{m,j}\mathcal{P}_t] + \text{Idle}_{N_{m,i}^c+1}\mathcal{P}_t + S_{m,i}\mathcal{P}_t$.

During AP duration, the BS broadcasts an announcement message to all active devices to announce the end of COP and the start of TOP. Hence, $L$ number of devices receive the message and let $E_{\text{AP}}$ denote the total energy used for receiving the announcement message from the BS during AP

$$E_{\text{AP}} = L\mathcal{P}_r T_{\text{ANC}}$$

where $T_{\text{ANC}}$ is the length of the announcement message.

During TOP duration, after receiving the allocated transmission schedule from the BS, each device sends its data packet to the BS at its scheduled time slots $T_r$. The energy consumption by successful devices in transmission during a single frame is defined as follows:

$$E_s = M_{\text{opt}}\mathcal{P}_t T_r.$$

The devices that failed in contention stay idle and keep their radio module OFF during TOP. Thus, over a single frame, it consumes the following energy:

$$E_{\text{in}} = (L - M_{\text{opt}})M_{\text{opt}}\mathcal{P}_i T_r$$

where $M_{\text{opt}}\mathcal{P}_i T_d$ denotes the energy used in idle mode during TOP. Hence, the total energy consumed by the devices during TOP is

$$E_{\text{TOP}} = E_s + E_{\text{in}}.$$

Therefore, the total energy consumption of all devices during one frame is defined as follows:

$$E_{\text{frame}} = E_{\text{NP}} + E_{\text{COP}} + E_{\text{AP}} + E_{\text{TOP}}. \qquad (23)$$

In simulation, we aim to compare the energy consumption in a frame among the proposed hybrid protocol, $p$-persistent CSMA and TDMA. The packet arrival rate of each device is $\lambda = 1$. For the proposed hybrid protocol, $p_{\text{inl}} = 0.1$ and $\alpha = 1$. Fig. 13 shows the comparison of energy consumption among the proposed hybrid method, *p*-persistent CSMA and TDMA. The comparison indicates that the proposed hybrid protocol is able to consume less energy than *p*-persistent CSMA. This is because the hybrid protocol only allows the devices to transmit a small length contending message during contention period. The energy consumption during collision can be greatly reduced. Moreover, when the contention is finished, the proposed hybrid method can control the devices that failed in contention turn to the idle mode for saving energy. In addition, the proposed protocol consumed more energy compared to that of TDMA scheme. That is, in the proposed protocol, devices have to use more energy for contending during the contention period; however, this energy consumption can lead to higher channel utility and lower packet drop ratio as shown in Sections V-A and V-B.

## VI. Conclusion

In this paper, we focused on designing the massive MAC protocol for heterogeneous M2M network where the devices have different service requirements. In our protocol, the operation of each frame is mainly divided into two parts: 1) COP and 2) TOP. The heterogeneous devices with different contending probability contend the transmission time slots during COP and only the successful devices in contention will be assigned the time slots for transmission. Considering the fairness, the contending probability of the device that failed in contention at previous frame will be increased at the next frame. Under such mechanism, the BS can easily maximize the channel utility by controlling the duration of COP $T_{\text{COP}}$, initial contending probability $p_{\text{inl}}$, and the incremental indicator $\alpha$. An optimization problem was formulated to solve the problem, and we showed analytically that the problem is convex. We analyzed the channel utility, packet drop ratio, average transmission delay, and energy consumption to show the effectiveness of the propose hybrid MAC protocol; especially, there are heterogeneous devices with different priorities.

## Appendix A
### Derivation of $U_\varrho^{(i-1)}$

$U_\varrho^{(i-1)}, i = 1, \ldots, I$ is defined as the number of empty class $\varrho$ devices that have new packet arrival during $(i-1)$th frame. Recall that the packet arrival process is a Possion arrival process with arrival rate $\lambda$ at each device. Let $g_\varrho$ denote the probability that a type $\varrho$ device has at least one new packet arrival during $T_{\text{frame}}$. Then, we have

$$g_\varrho = 1 - e^{-\lambda T_{\text{frame}}}.$$

Next, we first calculate $U_\varrho^{(i-1)}$ when $i = 1$. In this case, $U_\varrho^{(0)}$ denote the number of empty class $\varrho$ devices that have new packet arrival during frame 0. In frame 0, the vitual priorities of devices are only decided by the heterogeneous type in M2M networks; hence, $K_\varrho = K_q, \varrho, q = 1, \ldots, Q$. Let $N_\varrho^0$ represent the number of class $\varrho$ devices that have at least one new packet arrival during frame 0. We can obtain the probability that $N_\varrho^0 = n$ as

$$P\{N_\varrho^0 = n\} = \binom{K_q}{n}[1 - (1-g_\varrho)^{T_{\text{frame}}}]^n \cdot (1-g_\varrho)^{T_{\text{frame}}(K_q - n)}.$$

Therefore, we can obtain

$$U_\varrho^{(0)} = \sum_{n=1}^{K_q} n P\{N_\varrho^0 = n\}.$$

For $i = 2, \ldots, I$, the number of empty $\varrho$ type devices $K_\varrho^{(i-1)} = K_q - M_\varrho^{(i-1)}$, where $M_\varrho^{(i-1)} = \lceil M^{(i-1)} P_{\text{pck}}(\varrho) \rceil$ and $P_{\text{pck}}(\varrho)$ is given by (20). Let $N_\varrho^{(i-1)}$ denote the number of empty class $\varrho$ devices that have at least one new packet arrival during $(i-1)$th frame and $P\{N_\varrho^{(i-1)} = n\}$ denote the probability that $N_\varrho^{(i-1)} = n$, we have

$$P\{N_\varrho^{(i-1)} = n\} = \binom{K_\varrho^{(i-1)}}{n}[1 - (1-g_\varrho)^{T_{\text{frame}}}]^n \cdot (1-g_\varrho)^{T_{\text{frame}}(K_\varrho^{(i-1)} - n)}.$$

Similarly, we can calculate

$$U_\varrho^{(i-1)} = \sum_{n=1}^{K_\varrho^{(i-1)}} n P\{N_\varrho^{(i-1)} = n\}, \quad i = 2, \ldots, I.$$

## Appendix B
### Proof of Theorem 1

Since the duration of $T_{\text{frame}}$ has a finite value, as $L \to \infty$, it is easy to obtain $L \gg M^{(i)}$, then we have

$$\mathcal{T}_{\text{COP}}^{(i)}(M^{(i)}, \alpha, p_{\text{inl}}) = M^{(i)} \cdot \left\{ \frac{(1-(1+\alpha)p_{\text{inl}})^L}{L(1+\alpha)p_{\text{inl}}(1-(1+\alpha)p_{\text{inl}})^{L-1}} \cdot \delta_{\text{idle}} \right.$$
$$+ \left( \frac{1-(1-(1+\alpha)p_{\text{inl}})^L}{L(1+\alpha)p_{\text{inl}}(1-(1+\alpha)p_{\text{inl}})^{L-1}} - 1 \right) \delta_{\text{coll}}$$
$$\left. + \delta_{\text{succ}} \right\}. \quad (24)$$

Moreover, $(1-(1+\alpha)p_{\text{inl}})^{L-1}$ tends to $(1-(1+\alpha)p_{\text{inl}})^L$ if $L$ is sufficiently large. Hence, we can obtain the approximated transformation of the above-mentioned equation as

$$\mathcal{T}_{\text{COP}}^{(i)}(M^{(i)}, \alpha, p_{\text{inl}}) = M^{(i)} \left\{ \frac{1}{L(1+\alpha)p_{\text{inl}}} \delta_{\text{idle}} + \delta_{\text{succ}} \right.$$
$$+ \left( \frac{1}{L(1+\alpha)p_{\text{inl}}(1-(1+\alpha)p_{\text{inl}})^{L-1}} \right.$$
$$\left. \left. - \frac{1}{L(1+\alpha)p_{\text{inl}}} - 1 \right) \delta_{\text{coll}} \right\}. \quad (25)$$

Taking the second derivative of $\mathcal{T}_{\text{COP}}^{(i)}(M^{(i)}, \alpha, p_{\text{inl}})$ with respect to $M^{(i)}$, $\alpha$, and $p_{\text{inl}}$, respectively, the Hessian matrix is given by

$$\boldsymbol{H} = \begin{pmatrix} \frac{\partial^2 \mathcal{T}_{\text{COP}}^{(i)}}{\partial M^{(i)^2}} & \frac{\partial^2 \mathcal{T}_{\text{COP}}^{(i)}}{\partial M^{(i)} \partial p_{\text{inl}}} & \frac{\partial^2 \mathcal{T}_{\text{COP}}^{(i)}}{\partial M^{(i)} \partial \alpha} \\ \frac{\partial^2 \mathcal{T}_{\text{COP}}^{(i)}}{\partial p_{\text{inl}} \partial M^{(i)}} & \frac{\partial^2 \mathcal{T}_{\text{COP}}^{(i)}}{\partial p_{\text{inl}}^2} & \frac{\partial^2 \mathcal{T}_{\text{COP}}^{(i)}}{\partial p_{\text{inl}} \partial \alpha} \\ \frac{\partial^2 \mathcal{T}_{\text{COP}}^{(i)}}{\partial \alpha \partial M^{(i)}} & \frac{\partial^2 \mathcal{T}_{\text{COP}}^{(i)}}{\partial \alpha \partial p_{\text{inl}}} & \frac{\partial^2 \mathcal{T}_{\text{COP}}^{(i)}}{\partial \alpha^2} \end{pmatrix}.$$

Recall that $L \to \infty$, it is easy to obtain

$$\frac{\partial^2 \mathcal{T}_{\text{COP}}^{(i)}}{\partial M^{(i)^2}} = 0$$

$$\frac{\partial^2 \mathcal{T}_{\text{COP}}^{(i)}}{\partial p_{\text{inl}}^2} = \frac{2}{L(1+\alpha)p_{\text{inl}}^3}\delta_{\text{idle}}$$
$$+ \left( \frac{1 + (1-(1+\alpha)p_{\text{inl}})^{L+1} + L(1+\alpha)p_{\text{inl}}}{(1+\alpha)p_{\text{inl}}^2(1-(1+\alpha)p_{\text{inl}})^{L+2}} \right)\delta_{\text{coll}} > 0$$

$$\frac{\partial^2 \mathcal{T}_{\text{COP}}^{(i)}}{\partial \alpha^2} = \frac{2}{Lp_{\text{inl}}(1+\alpha)^3}\delta_{\text{idle}}$$
$$+ \left( \frac{1 + (1-(1+\alpha)p_{\text{inl}})^{L+1} + L(1+\alpha)p_{\text{inl}}}{(1+\alpha)^2 p_{\text{inl}}(1-(1+\alpha)p_{\text{inl}})^{L+2}} \right)\delta_{\text{coll}} > 0$$

$$\frac{\partial^2 \mathcal{T}_{\text{COP}}^{(i)}}{\partial M^{(i)} \partial p_{\text{inl}}} = \frac{\partial^2 \mathcal{T}_{\text{COP}}^{(i)}}{\partial p_{\text{inl}} \partial M^{(i)}} = \frac{-1}{L(1+\alpha)p_{\text{inl}}^2}\delta_{\text{idle}}$$
$$+ \left( \frac{(1-(1+\alpha)p_{\text{inl}})^L + (L-1)(1+\alpha)^2 p_{\text{inl}}}{(L-1)(1+\alpha)p_{\text{inl}}(1-(1+\alpha)p_{\text{inl}})^{L+1}} \right)$$
$$\times \delta_{\text{coll}} > 0$$

$$\frac{\partial^2 \mathcal{T}_{\text{COP}}^{(i)}}{\partial M^{(i)} \partial \alpha} = \frac{\partial^2 \mathcal{T}_{\text{COP}}^{(i)}}{\partial M^{(i)} \partial \alpha} = \frac{-1}{L(1+\alpha)^2 p_{\text{inl}}}\delta_{\text{idle}}$$
$$+ \left( \frac{(1-(1+\alpha)p_{\text{inl}})^{L-1} + L(1+\alpha)p_{\text{inl}}^2}{L(1+\alpha)p_{\text{inl}}(1-(1+\alpha)p_{\text{inl}})^{L+1}} \right)\delta_{\text{coll}} > 0$$

$$\frac{\partial^2 \mathcal{T}_{\text{COP}}^{(i)}}{\partial M^{(i)} \partial p_{\text{inl}}} = \frac{\partial^2 \mathcal{T}_{\text{COP}}^{(i)}}{\partial p_{\text{inl}} \partial M^{(i)}} = \frac{2}{L(1+\alpha)^2 p_{\text{inl}}^2}\delta_{\text{idle}}$$
$$+ \left( \frac{(1-(1+\alpha)p_{\text{inl}})^L + L(1+\alpha)p_{\text{inl}}}{(1+\alpha)p_{\text{inl}}(1-(1+\alpha)p_{\text{inl}})^{L+2}} \right)\delta_{\text{coll}} > 0.$$

Consequently, the Hessian matrix of $\mathcal{T}_{\text{COP}}$, $\boldsymbol{H} \geq 0$, we conclude that $\mathcal{T}_{\text{COP}}^{(i)}(M^{(i)}, \alpha, p_{\text{inl}})$ is a convex function of $M^{(i)}$, $\alpha$, and $p_{\text{inl}}$ [30]. ∎

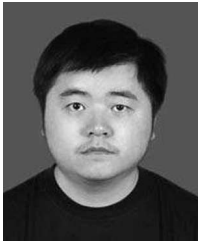

**Yi Liu** received the Ph.D. degree from South China University of Technology (SCUT), Guangzhou, China, in 2011.

After that, he worked with the Institute of Intelligent Information Processing, Guangdong University of Technology (GDUT), Guangzhou, China. In 2011, he joined as a Postdoctoral with the Singapore University of Technology and Design, Singapore. His research interests include cognitive radio networks, cooperative communications, smart grid and intelligent signal processing.

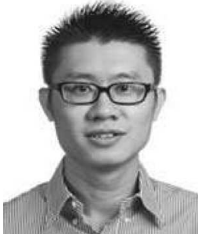

**Chau Yuen** (S'02–M'08–SM'12) received the B.Eng. and Ph.D. degrees from Nanyang Technological University, Singapore, in 2000 and 2004, respectively.

He was a PostDoc Fellow at Lucent Technologies Bell Labs (Murray Hill) during 2005, and a Visiting Assistant Professor with Hong Kong Polytechnic University, Hong Kong, in 2008. From 2006 to 2010, he worked as a Senior Research Engineer with the Institute for Infocomm Research, Singapore. He joined as an Assistant Professor with Singapore University of Technology and Design in June 2010. He has published over 150 research papers in international journals or conferences. His current research interests include green communications, massive MIMO, Internet-of-things, machine-to-machine, network coding, and distributed storage.

Dr. Yuen also serves as an Associate Editor for the IEEE TRANSACTIONS ON VEHICULAR TECHNOLOGY.

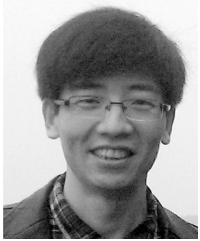

**Xianghui Cao** (S'08–M'11) received the B.S. and Ph.D. degrees in control science and engineering from Zhejiang University, Hangzhou, China, in 2006 and 2011, respectively.

From 2007 to 2009, he was a visiting scholar with the Department of Computer Science, University of Alabama, Tuscaloosa, USA. He is an Associate Editor for KSII *Transactions on Internet and Information Systems and Security and Communication Networks* (Wiley). His research interests include wireless network performance analysis, energy efficiency of wireless networks, networked estimation and control, and network security.

Dr. Cao is a TPC member for IEEE Globecom 2013, 2014, IEEE ICC 2014, IEEE VTC 2013, 2014, etc.

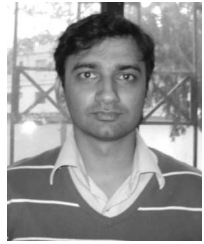

**Naveed Ul Hassan** received the B.E. degree in avionics engineering from the College of Aeronautical Engineering, Risalpur, Pakistan, in 2002, the Master's and Ph.D. degrees in telecommunications from Ecole Superieure d'Electricite (Supelec), Gif-sur-Yvette, France, 2006 and 2010, respectively.

Since August 2011, he serves as an Assistant Professor with the Department of Electrical Engineering, Lahore University of Management Sciences (LUMS), Lahore, Pakistan. His research interests include cross layer design and optimization in wireless networks, heterogeneous networks, cognitive radio networks and smart grids.

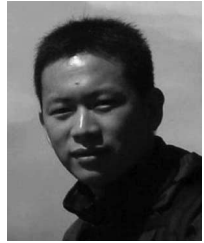

**Jiming Chen** (M'08–SM'11) received the B.Sc. and Ph.D. degrees both in control science and engineering from Zhejiang University, China, in 2000 and 2005, respectively.

He was a Visiting Researcher with Institut National de Recherche en Informatique et en Automatique (INRIA), France, in 2006, National University of Singapore, Singapore, in 2007, and University of Waterloo, Waterloo, QC, Canada, from 2008 to 2010. Currently, he is a Full Professor with the Department of Control Science and Engineering, the Coordinator of Group of Networked Sensing and Control in the State Key Laboratory of Industrial Control Technology, and Vice Director of the Institute of Industrial Process Control, Zhejiang University, Zhejiang, China.

Dr. Chen currently serves as an Associate Editor for several international journals including IEEE TRANSACTIONS ON PARALLEL AND DISTRIBUTED SYSTEM, IEEE TRANSACTIONS ON INDUSTRIAL ELECTRONICS, IEEE NETWORK, IEEE TRANSACTION ON CONTROL OF NETWORK SYSTEMS, etc. He was a Guest Editor of IEEE TRANSACTIONS ON AUTOMATIC CONTROL, *Computer Communication* (Elsevier), *Wireless Communication and Mobile Computer* (Wiley) and *Journal of Network and Computer Applications* (Elsevier). He also served/serves as Ad hoc and Sensor Network Symposium Co-chair, IEEE Globecom 2011; general symposia Co-Chair of ACM IWCMC 2009 and ACM IWCMC 2010, WiCON 2010 MAC track Co-Chair, IEEE MASS 2011 Publicity Co-Chair, IEEE DCOSS 2011 Publicity Co-Chair, IEEE ICDCS 2012 Publicity Co-Chair, IEEE ICCC 2012 Communications QoS and Reliability Symposium Co-Chair, IEEE SmartGridComm The Whole Picture Symposium Co-Chair, IEEE MASS 2013 Local Chair, Wireless Networking and Applications Symposium Co-chair, IEEE ICCC 2013, Ad hoc and Sensor Network Symposium Co-chair, IEEE ICC 2014, and TPC member for IEEE ICDCS'10,'12,'13,'14, IEEE MASS'10,'11,'13, IEEE SECON'11,'12, IEEE INFOCOM'11,'12,'13,'14, etc.